\documentclass[prd,aps,showpacs,tightenlines,nofootinbib,preprint,preprintnumbers]{revtex4}
\usepackage{graphicx}
\usepackage{amsfonts}
\usepackage{amssymb}
\usepackage{latexsym}
\usepackage{color}
\input{colordvi.tex}

\newcommand{\CO}{{\cal O}}

\newcommand{\bear}{\begin{array}}  \newcommand{\eear}{\end{array}}
\newcommand{\bea}{\begin{eqnarray}}  \newcommand{\eea}{\end{eqnarray}}
\newcommand{\beq}{\begin{equation}}  \newcommand{\eeq}{\end{equation}}
\newcommand{\bef}{\begin{figure}}  \newcommand{\eef}{\end{figure}}
\newcommand{\bec}{\begin{center}}  \newcommand{\eec}{\end{center}}

\newcommand{\bib}{\bibitem}

\def\IBB#1#2#3{{\bf #1}, #2 (20#3)}

\def\APJSS#1#2#3{Astrophys. J. Suppl. {\bf #1}, L#2 (20#3)}

\def\CQGG#1#2#3{Class. Quantum Grav. {\bf #1}, #2 (20#3)}

\def\JCAPP#1#2#3{J. Cosmol. Astropart. Phys. {\bf #1}, #2 (20#3)}

\def\JHEPP#1#2#3{J. High Energy Phys. {\bf #1}, #2 (20#3)}

\def\NPB#1#2#3{Nucl. Phys. {\bf B#1}, #2 (19#3)}
\def\NPBB#1#2#3{Nucl. Phys. {\bf B#1}, #2 (20#3)}

\def\PLB#1#2#3{Phys. Lett. B {\bf #1}, #2 (19#3)}
\def\PLBB#1#2#3{Phys. Lett. B {\bf #1}, #2 (20#3)}

\def\PRD#1#2#3{Phys. Rev. D {\bf #1}, #2 (19#3)}
\def\PRDD#1#2#3{Phys. Rev. D {\bf #1}, #2 (20#3)}

\def\PRLL#1#2#3{Phys. Rev. Lett. {\bf#1}, #2 (20#3)}

\def\PRT#1#2#3{Phys. Rep. {\bf#1}, #2 (19#3)}

\begin{document}


\title{Non-Gaussianity in the modulated reheating scenario}
\author{Teruaki Suyama}
\affiliation{Institute for Cosmic Ray Research, University of Tokyo,
Kashiwa 277-8582, Japan}
\author{Masahide Yamaguchi} 
\affiliation{Department of Physics and Mathematics, Aoyama Gakuin
University, Sagamihara 229-8558, Japan}
\date{\today}
%
%
\begin{abstract}
We investigate the non-Gaussianity of primordial curvature perturbations
in the modulated reheating scenario where the primordial perturbation
is generated due to the spatial fluctuation of the rate of the inflaton 
decay to radiation.
We use the $\delta N$ formalism to
evaluate the trispectrum of the curvature perturbation as well as its
bispectrum.  We give expressions for three non-linear parameters
$f_{NL},~\tau_{NL}$ and $g_{NL}$ in the modulated reheating scenario.
If both the intrinsic non-Gaussianity of scalar field fluctuations and third
the derivative of the decay rate with respect to the scalar fields are negligibly small, 
$g_{NL}$ has at least the same order of magnitude as $f_{NL}$.  
We also give a general inequality between $f_{NL}$ and $\tau_{NL}$, 
which is true for other inflationary scenarios as long as the 
primordial non-Gaussianity comes from super-horizon evolution.
\end{abstract}


\maketitle


\section{Introduction}

Recent observations of cosmic microwave background (CMB) anisotropies
give strong evidence that primordial density fluctuations are almost
Gaussian, scale-invariant, and adiabatic \cite{WMAP3}. 
During inflation,
they are generated as vacuum fluctuations of light fields and are
stretched to cosmological scales to explain the large scale structure 
of the universe \cite{inflation}. 
Such a light field responsible for density fluctuations 
has been considered to be the inflaton itself for a long time.

Recently, alternative candidates for such a light field have been
proposed. 
One attractive example is the curvaton \cite{precur,curvaton},
which is effectively massless and acquires fluctuations during
inflation. 
After inflation, it becomes effectively massive and
contributes to a non-negligible fraction of the energy density of the
universe. 
Then, 
after it decays, eventually density fluctuations induced
by the curvaton are converted to adiabatic ones and can dominate over
those generated by the inflaton itself. 
Another interesting candidate is a light field whose expectation 
value determines the coupling constant of the inflaton to standard 
model particles \cite{modulated}. 
Such a light field will fluctuate during inflation, 
which leads to fluctuation of the decay rate of the inflaton. 
Then, 
this latter spatial fluctuation induces that of the reheating temperature,
which eventually generates curvature perturbations. 
Some variants of these two alternative candidates have
also been considered related to the fluctuations of masses and
annihilation cross sections \cite{masscross}, and the preheating
mechanism \cite{preheating}.

In order to determine which light field is actually responsible for
primordial density fluctuations, 
the deviation from Gaussianity of the curvature perturbations is of great use. 
If density fluctuations are completely Gaussian, their bispectra, 
characterized by a parameter $f_{NL}$, 
and (connected) trispectra, 
characterized by parameters $g_{NL}$ and $\tau_{NL}$ vanish.  
Therefore, estimation of the bispectra and trispectra
is important to identify the light field. 
In \cite{malda}, 
it was shown that the bispectrum is significantly suppressed by 
the slow-roll parameters up to an undetectable level in a
single field slow-roll inflation model.\footnote{It has recently
pointed out that there is a possibility that an all-sky 21-cm experiment is
sensitive to a value of $|f_{NL}| \sim 0.01$ \cite{cooray}.} 
After \cite{malda},
the trispectrum, 
as well as bispectrum, 
were evaluated in a multi-field configuration and a curvaton scenario. 
Though spectra are still suppressed by slow-roll parameters in a multi-field 
configuration \cite{multi,Alabidi:2006wa}, 
they can be significantly large in the curvaton scenario \cite{lcurvaton,Byrnes:2006vq}. 
On the contrary, 
only the bispectrum is calculated in the modulated reheating scenario
\cite{zal,BMR,Vernizzi:2003vs}.  
Since the bispectrum can be large in both the curvaton scenario and 
the modulated reheating scenario, 
trispectrum may be useful to discriminate between them. 
Though the present constraint on the trispectrum is not particularly severe, 
and is roughly given by $|\tau_{NL}| \lesssim 10^{8}$ \cite{Alabidi:2005qi}, 
a value of $|\tau_{NL}| \sim 560$ will be detectable by the Planck satellite \cite{KK}.
\footnote{In order to
obtain these constraints, one must compute not only the
non-Gaussianities of primordial fluctuations but also other non-linear
effects entering in the CMB such the non-linear evolution of the
perturbations after inflation \cite{CMBnon}.}

The main purpose of our paper is to estimate the trispectrum in the
modulated reheating scenario. 
We use the $\delta N$ formalism,
which is a powerful approach to evaluate the non-Gaussianity 
of curvature perturbations simply because it requires a 
homogeneous background solution \cite{Starobinsky:1986fx, Sasaki:1995aw, Sasaki:1998ug, Lyth:2004gb}.
Then, we use the same formalism to evaluate the trispectrum of
curvature perturbations as well as its bispectrum.

This paper is organized as follows. 
In the next section, 
we give a brief review of the $\delta N$ formalism
and a definition of the three non-linear parameters $f_{NL},~\tau_{NL}$ 
and $g_{NL}$ given in \cite{Lyth:2005fi,Alabidi:2005qi,Byrnes:2006vq}.
We also give a general inequality between $f_{NL}$ and
$\tau_{NL}$ which is not found in the literature.
In Sec.~III,
we study the background dynamics in the modulated reheating scenario.
In Sec. IV, 
we study the perturbations in this scenario and give expressions 
for $f_{NL},~\tau_{NL}$ and $g_{NL}$.
The final section is devoted to a summary.
We use the units $8\pi G=1$.

\section{$\delta N$ formalism}

According to the $\delta N$ formalism \cite{Starobinsky:1986fx, Sasaki:1995aw, Sasaki:1998ug, Lyth:2004gb},
the curvature perturbation on a uniform energy 
density hypersurface $\zeta$ at time $t_f$ is,
on sufficiently large scales,
equal to the perturbation in the time integral of the local 
expansion from an initial flat hypersurface ($t=t_i$) to 
the final uniform energy density hypersurface.
On sufficiently large scales,
the local expansion can be approximated quite well 
by the expansion of the unperturbed Friedmann universe.
Hence
\begin{eqnarray}
\zeta (t_f, {\vec x}) = N(t_i,t_f,{\vec x})-({\rm spatial~average}), \label{delN1}
\end{eqnarray}
where the $e$-folding number $N(t_i,t_f,{\vec x})$ is defined by
the time integral of the local Hubble parameter,
\begin{eqnarray}
N(t_i,t_f,{\vec x})=\int_{t_i}^{t_f} H(t,{\vec x})dt. \label{delN2}
\end{eqnarray}

In many inflationary scenarios, 
which include the modulated reheating scenario,
the dynamics of the universe between $t_i$ and $t_f$
is determined by the values of the relevant scalar fields $\phi^I$ 
at $t_i$ and by the $e$-folding number which becomes a function
of $\phi^I(t_i,{\vec x})$.
Hence the curvature perturbation at $t_f$ is given by
\begin{eqnarray}
\zeta (t_f,{\vec x}) \approx N_I \delta \phi^I+\frac{1}{2} N_{IJ} \delta \phi^I \delta \phi^J+\cdots-({\rm spatial~average}), \label{delN3}
\end{eqnarray}
where $\delta \phi^I$ is the perturbation of $\phi^I$
on the flat hypersurface at $t_i$, 
and $N_I,~N_{IJ},\cdots$ are given by
\begin{eqnarray}
N_I = \frac{\partial N}{\partial \phi^I},~~~N_{IJ}=\frac{\partial^2 N}{\partial \phi^I \partial \phi^J}, ~~~\cdots. \label{delN4}
\end{eqnarray}
Because solutions of the unperturbed Friedmann equation give
$N_I,N_{IJ},\cdots$,
the knowledge of the background solutions is enough to 
know the higher order correlation functions of $\zeta$.

The connected parts of the power spectrum, 
bispectrum and trispectrum are defined as
\begin{eqnarray}
&&\langle \zeta_{\vec k_1} \zeta_{\vec k_2} \rangle_c ={(2\pi)}^3 P_\zeta (k_1) \delta ({\vec k_1}+{\vec k_2}), \label{delN5} \\
&&\langle \zeta_{\vec k_1} \zeta_{\vec k_2} \zeta_{\vec k_3} \rangle_c={(2\pi)}^3 B_\zeta (k_1,k_2,k_3) \delta ({\vec k_1}+{\vec k_2}+{\vec k_3}), \label{delN6} \\
&&\langle \zeta_{\vec k_1} \zeta_{\vec k_2} \zeta_{\vec k_3} \zeta_{\vec k_4} \rangle_c ={(2\pi)}^3 T_\zeta (k_1,k_2,k_3,k_4) \delta ({\vec k_1}+{\vec k_2}+{\vec k_3}+{\vec k_4}). \label{delN7}
\end{eqnarray}
Here $\langle \cdots \rangle_c$ means that we take the connected
part of $\langle \cdots \rangle$.
Using Eq.~(\ref{delN3}),
we can express $P_\zeta,~B_\zeta$ and $T_\zeta$ in terms of the 
correlation functions of $\delta \phi^I$, which are given 
to leading order by \cite{Byrnes:2006vq},
\begin{eqnarray}
&&P_\zeta (k)=N_I N_J P^{IJ} (k), \label{delN8} \\
&&B_\zeta (k_1,k_2,k_3)=N_I N_J N_K B^{IJK} (k_1,k_2,k_3)+N_I N_{JK} N_K \left( P^{IK}(k_1) P^{JL} (k_2)+ {\rm 2~perms} \right), \label{delN9} \\
&&T_\zeta (k_1,k_2,k_3,k_4)=N_I N_J N_K N_L T^{IJKL}(k_1,k_2,k_3,k_4)\nonumber \\
&&\hspace{35mm}+N_{IJ}N_K N_L N_M \left( P^{IK}(k_1) B^{JLM} (k_{12},k_3,k_4)+11~{\rm perms.} \right) \nonumber \\
&&\hspace{35mm}+N_{IJ}N_{KL} N_M N_N \left( P^{JL}(k_{13})P^{IM}(k_3)P^{JN}(k_4)+11~{\rm perms.} \right) \nonumber \\
&&\hspace{35mm}+N_{IJK}N_L N_M N_N \left( P^{IL}(k_2)P^{JM}(k_3) P^{KN}(k_4)+3~{\rm perms.} \right), \label{delN10}
\end{eqnarray}
where $k_{ij}=|{\vec k_i}-{\vec k_j}|$ and 
$P^{IJ},~B^{IJK},~T^{IJKL}$ are the power spectrum,
bispectrum and trispectrum of the scalar fields respectively,
defined by
\begin{eqnarray}
&&\langle \delta \phi^I_{\vec k_1} \delta \phi^J_{\vec k_2} \rangle_c={(2\pi)}^3 P^{IJ} (k_1) \delta ({\vec k_1}+{\vec k_2}), \label{delN11} \\
&&\langle \delta \phi^I_{\vec k_1} \delta \phi^J_{\vec k_2} \delta \phi^K_{\vec k_3} \rangle_c={(2\pi)}^3 B^{IJK} (k_1,k_2,k_3) \delta ({\vec k_1}+{\vec k_2}+{\vec k_3}), \label{delN12} \\
&&\langle \delta \phi^I_{\vec k_1} \delta \phi^J_{\vec k_2} \delta \phi^K_{\vec k_3} \delta \phi^L_{\vec k_4} \rangle_c ={(2\pi)}^3 T^{IJKL} (k_1,k_2,k_3,k_4) \delta ({\vec k_1}+{\vec k_2}+{\vec k_3}+{\vec k_4}). \label{delN13}
\end{eqnarray}
If $\delta \phi^I$ are independent Gaussian variables with 
the same variance which we denote as $P$,
then $P^{IJ}=P \delta^{IJ}$ and $B^{IJK}$ and $T^{IJKL}$ vanish.
In such a case,
deviation from Gaussianity of the primordial perturbation 
comes only from super-horizon evolution, 
and the non-Gaussianity is characterized by three constant parameters 
$f_{NL},~\tau_{NL}$ and $g_{NL}$ defined by
\begin{eqnarray}
&&B_\zeta (k_1,k_2,k_3)=\frac{6}{5} f_{NL} \left( P_\zeta (k_1) P_\zeta (k_2)+2~{\rm perms.} \right), \label{delN14} \\
&&T_\zeta (k_1,k_2,k_3,k_4)=\tau_{NL} \left( P_\zeta(k_{13}) P_\zeta (k_3) P_\zeta (k_4)+11~{\rm perms.} \right) \nonumber \\
&&\hspace{35mm}+\frac{54}{25} g_{NL} \left( P_\zeta (k_2) P_\zeta (k_3) P_\zeta (k_4)+3~{\rm perms.} \right). \label{delN15} 
\end{eqnarray}
Using Eqs.~(\ref{delN8})-(\ref{delN10}),
we have the following expressions,
\begin{eqnarray}
&&f_{NL}=\frac{5}{6} \frac{N_I N_J N^{IJ}}{ {(N_K N^K)}^2 }, \label{delN16} \\
&&\tau_{NL}=\frac{N_{IJ} N^{IK} N^J N_K}{ {(N_L N^L)}^3 }, \label{delN17} \\
&&g_{NL}=\frac{25}{54} \frac{N_{IJK} N^I N^J N^K}{ {(N_L N^L)}^3 }. \label{delN18}
\end{eqnarray}
Eq.~(\ref{delN16}) has been given in \cite{Lyth:2005fi}.
Eq.~(\ref{delN17}) has been given in the arXiv version of \cite{Alabidi:2005qi}
and also in \cite{Byrnes:2006vq}.
Eq.~(\ref{delN18}) has been given in \cite{Byrnes:2006vq}.

Here we provide a relation between $f_{NL}$ and $\tau_{NL}$ which is not
found in the literature. 
From the Cauchy-Schwarz inequality, 
we have the following
\begin{eqnarray}
\tau_{NL} \ge \frac{36}{25} f_{NL}^2. \label{delN19}
\end{eqnarray}
We have equality if and only if the vector $N_I$ is an eigenvector of
the matrix $N_{IJ}$.  
The single inflation model yields $\tau_{NL}=\frac{36}{25} f_{NL}^2$.  
However in multi-field inflation,
there is a possibility that two vectors $N_I$ and $N_{IJ}N^J$ are nearly
orthogonal.  
In such a case, $f_{NL}$ is very small but $\tau_{NL}$ and
possibly also $g_{NL}$ remains finite.
Hence the leading non-Gaussianity comes not from the bispectrum 
but from the trispectrum.

\section{Background dynamics of the modulated reheating scenario}

In the modulated reheating scenario \cite{modulated}, the decay rate
$\Gamma$ of the inflaton $S$ is a function of scalar fields $\phi^I$
(not necessarily a single field) which are light during inflation.  We
assume that fluctuation of the inflaton field generates negligible
curvature perturbation.  Then the detailed form of the inflaton
potential $U(S)$ during inflation is not important for the scenario.  We
only require that $U(S)$ around the minimum is approximated well by a
term quadratic in $S$.  
After inflation, the inflaton oscillates around the
minimum of the potential.  The energy density of the inflaton $\rho_S$
averaged over one period of the oscillation behaves as a function of
the $e$-folding number $\propto e^{-3N}$. 
Hence we regard $\rho_S$ as dust.  
The inflaton decays into radiation with rate $\Gamma$ which depends 
on the expectation values of $\phi^I$.  
Then the background equations are given by
\begin{eqnarray}
&&\frac{d \rho_S}{dN}+3 \rho_S=-\frac{\Gamma}{H} \rho_S, \label{bd1} \\
&&\frac{d \rho_r}{dN}+4 \rho_r=\frac{\Gamma}{H} \rho_S, \label{bd2} \\
&&H^2=\frac{1}{3} (\rho_S+\rho_r), \label{bd3}
\end{eqnarray}
where $\rho_r$ is the energy density of radiation.  
Spatial fluctuations of these light fields induce the fluctuation of 
the decay rate and the curvature perturbation. 
By solving the above equations from the end of
inflation to the completion of reheating with the initial conditions
$\rho_S(0)=\rho_0=3H_0^2,~\rho_r(0)=0$, we can obtain a relation between
$N$ and $\Gamma$. 
Until the Hubble function drops to $H_f$ ($H_f \ll \Gamma$),
the $e$-folding number $N$ can be written formally as
\begin{eqnarray}
N =\frac{1}{2} \log \frac{H_0}{H_f}+Q \left( \frac{\Gamma}{H_0} \right), \label{bd4}
\end{eqnarray}
where
\begin{eqnarray}
\exp \bigg[ 4Q ( \Gamma /H_0 ) \bigg] \equiv \int_0^\infty dN'~\frac{\Gamma}{H(N)}e^{4N'}\frac{\rho_S(N')}{\rho_0}. \label{bd5}
\end{eqnarray}

\begin{figure}[t]
  \includegraphics[width=9cm,clip]{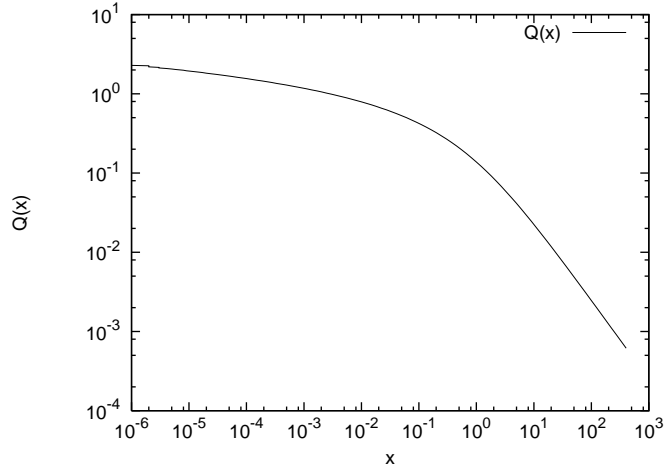}
\caption{Plot of $Q(x)$ calculated numerically.
The fitting formula $Q_{\rm fit}(x)$ lies within the
solid line. 
}
\label{fig1} 
\end{figure}

For the two limiting cases,
we have the approximate form of $Q(x)$ as
\begin{eqnarray}
&&Q(x) =\frac{1}{4x}+{\cal O}(x^{-2})~~~~~x\gg 1, \label{bd6} \\
&&Q(x)=-\frac{1}{6} \log x+{\cal O}(x) ~~~~~x\ll 1. \label{bd7}
\end{eqnarray}
For arbitrary $x$,
we do not have an analytic form for $Q(x)$ and
we have to solve the background equations numerically.
We show $Q(x)$ calculated numerically in Fig.~\ref{fig1}.
We also find an accurate fitting formula for $Q(x)$
of the form,
\begin{eqnarray}
Q_{\rm fit}(x)=\frac{1}{4} ~\frac{r(x)+2}{r(x)+3} \log \left( 1+\frac{1}{x} \right). \label{bd8}
\end{eqnarray}
For $r(x)=2.16 x^{0.72}$,
the relative error is within 2 percent.
For $r(x)=1.7 x^{0.9}+0.3 x^{0.18}$,
the relative error is within $0.5$ percent.

\section{Perturbation in the modulated reheating scenario}

From Eq.~(\ref{bd4}),
we can calculate $\delta N$ in terms of $\delta \Gamma$,
\begin{eqnarray}
\delta N=xQ'(x) \frac{\delta \Gamma}{\Gamma}+\frac{x^2}{2}Q''(x) {\left(
								   \frac{\delta
								   \Gamma}{\Gamma}
								  \right)}^2+\frac{x^3}{6} Q'''(x) {\left( \frac{\delta \Gamma}{\Gamma} \right)}^3 + \cdots. \label{pm1}
\end{eqnarray}
In the modulated reheating scenario,
the decay rate depends on the scalar fields which
are almost massless during inflation.
Hence the perturbation of the decay rate can be written
as a function of the perturbation of the scalar fields,
\begin{eqnarray}
\delta \Gamma=\Gamma_I \delta \phi^I+\frac{1}{2} \Gamma_{IJ} \delta
 \phi^I \delta \phi^J+\frac{1}{6} \Gamma_{IJK} \delta \phi^I \delta
 \phi^J \delta \phi^K + \cdots. \label{pm2}
\end{eqnarray}
From these equations,
$N_I,~N_{IJ},~N_{IJK}$ can be written as
\begin{eqnarray}
&&N_I=xQ'(x) \frac{\Gamma_I}{\Gamma}, \label{pm3} \\
&&N_{IJ}=xQ'(x) \frac{\Gamma_{IJ}}{\Gamma}+x^2Q''(x) \frac{\Gamma_I}{\Gamma} \frac{\Gamma_J}{\Gamma}, \label{pm4} \\
&&N_{IJK}=xQ'(x) \frac{\Gamma_{IJK}}{\Gamma}+x^2 Q''(x) \left( \frac{\Gamma_I}{\Gamma} \frac{\Gamma_{JK}}{\Gamma}+\frac{\Gamma_J}{\Gamma} \frac{\Gamma_{KI}}{\Gamma}+\frac{\Gamma_K}{\Gamma} \frac{\Gamma_{IJ}}{\Gamma} \right)+x^3 Q'''(x) \frac{\Gamma_I}{\Gamma} \frac{\Gamma_J}{\Gamma} \frac{\Gamma_K}{\Gamma}. \label{pm5}
\end{eqnarray}
Substituting these latter into Eqs.~(\ref{delN8})-(\ref{delN10})
yields the power spectrum, bispectrum and trispectrum of the primordial
curvature perturbation in the modulated reheating scenario. Note that
Eqs.~(\ref{pm3})-(\ref{pm5}) are only correct if $t_i$ is taken as the
time when inflation ends. 
Hence in Eqs.~(\ref{delN8})-(\ref{delN10}), 
we must use the power spectrum, bispectrum and
trispectrum of scalar field fluctuation at the end of inflation.

Zaldarriaga \cite{zal} evaluated the bispectrum $B$ of the curvature
perturbations at reheating by taking into account not only the nonlinear
evolution of the scalar fluctuations but also their intrinsic
non-Gaussianities \cite{zal}. In particular, in order to estimate the
intrinsic non-Gaussianity of a scalar field, he calculates the leading
order quantum three-point correlation function with the assumption that
$\Gamma$ depends only on a single field $\phi$,
\begin{eqnarray}
\langle {\hat \phi}_{\vec k_1}(\eta) {\hat \phi}_{\vec k_2}(\eta) {\hat \phi}_{\vec k_3}(\eta) \rangle_c=-i \langle 0| \int_{-\infty}^\eta d\eta'~\big[{\hat \phi}_{\vec k_1} (\eta) {\hat \phi}_{\vec k_2}(\eta) {\hat \phi}_{\vec k_3}(\eta),{\hat V}(\eta') \big] 0\rangle, \label{pm6}
\end{eqnarray}
where $\eta \equiv \int_{-\infty}^t \frac{dt}{a}$ is the conformal
time. 
In Ref.~\cite{zal}, mode functions in pure de-Sitter spacetime are used
throughout the evolution of the above correlation function, though the use of
such mode functions are not necessarily justified during the oscillatory
phase after inflation. 
In our formalism, 
we have only to know the field fluctuations at the end of inflation. 
However, even in this case, 
mode functions in pure de-Sitter spacetime are not good approximations 
after the relevant scale crosses the horizon.  
This is because the Hubble parameter may decrease by an order of magnitude 
from the horizon exit to the end of inflation. 
In this paper, instead of using Eq.~(\ref{pm6}) until the end of inflation, 
we use it only until the relevant mode exits the horizon. 
After the horizon crossing, we evolve $\delta \phi^I$ by perturbing the 
classical unperturbed equations, 
in keeping with the spirit of the $\delta N$ formalism, which correctly
takes into account the evolution of the Hubble parameter.

We start from the evolution of scalar field fluctuations
on super-horizon scales,
which can be described well by the classical treatment.
Then, the background equations are given by
\begin{eqnarray}
\frac{d^2 \phi^I}{dN^2}+\left( 3+\frac{1}{H} \frac{dH}{dN} \right) \frac{d\phi^I}{dN}+\frac{V^I}{H^2}=0. \label{pm7}
\end{eqnarray}
We assume that scalar fields slow-roll during the whole epochs of interest,
which enables us to approximate the background equations as
\begin{eqnarray}
\frac{d\phi^I}{dN} \simeq -\frac{1}{3+\frac{H'}{H}} \frac{V^I}{H^2}. \label{pm8}
\end{eqnarray}
Throughout this paper,
it is also assumed that the total energy density of the universe is
dominated by the inflaton potential, 
and the dependence of $\phi^I$ on $H$ is negligible.

Let $N_*$ be the $e$-folding slightly after the horizon crossing and
$N_f$ be the $e$-folding at the end of inflation. 
Slightly after the horizon crossing, 
the scalar field fluctuations turn into classical variables with their 
magnitude given by $\delta \phi^I_*$.  
Then $\delta \phi^I (N)$ after $N_*$ is a function of $\delta \phi^I_*$, 
which can be Taylor-expanded with respect to $\delta \phi^I_*$.  
For the sake of the evaluation of the leading bispectrum and trispectrum 
of scalar fields at $N=N_f$, 
it is enough to expand $\delta \phi^I(N)$ to third order in
$\delta \phi^I_*$.  
Up to third order, Eq.~(\ref{pm8}) yields
\begin{eqnarray}
\phi^I(N_f)=\Lambda^I_J(N_f,N_\ast) \delta \phi^J_\ast+\frac{1}{2}\Theta^I_{JK} (N_f,N_\ast) \delta \phi^J_\ast \delta \phi^K_\ast+\frac{1}{6} \Xi^I_{JKL}(N_f,N_\ast) \delta \phi^J_\ast \delta \phi^K_\ast \delta \phi^L_\ast, \label{pm9}
\end{eqnarray}
where
\begin{eqnarray}
&&\Lambda^I_J(N,N') \equiv \bigg[ T \exp \left( \int^N_{N'} dN''~P(N'') \right) \bigg]^I_J, \label{pm10} \\
&&\Theta^I_{JK}(N_f,N_\ast) = \int_{N_\ast}^{N_f} dN' \Lambda^I_L (N_f,N') Q^L_{MN}(N') \Lambda^M_J (N',N_\ast) \Lambda^N_K(N',N_\ast),\label{pm11} \\
&&\Xi^I_{JKL}(N_f,N_\ast) =\frac{3}{2} \int_{N_\ast}^{N_f} dN'~\Lambda^I_{I'}(N,N')Q^{I'}_{MN}(N') \Lambda^M_J (N',N_\ast) \nonumber \\
&&\hspace{27mm}\times \int_{N_\ast}^{N'}dN''~\Lambda^N_{N'}(N',N'')Q^{N'}_{PQ}(N'')\Lambda^P_K(N'',N_\ast) \Lambda^Q_L (N'',N_\ast) \nonumber \\
&&\hspace{27mm}+\int_{N_\ast}^N dN'~\Lambda^I_{I'} (N,N') R^{I'}_{J'K'L'}(N') \Lambda^{J'}_J(N',N_\ast) \Lambda^{K'}_K (N',N_\ast) \Lambda^{L'}_L (N',N_\ast),\label{pm12} \\
&&P^I_J=-\frac{1}{3+\frac{H'}{H}} \frac{V^I_J}{H^2}, \label{pm13}\\
&&Q^I_{JK}=-\frac{1}{3+\frac{H'}{H}} \frac{V^I_{JK}}{H^2}, \label{pm14}\\
&&R^I_{JKL}=-\frac{1}{3+\frac{H'}{H}} \frac{V^I_{JKL}}{H^2}. \label{pm15}
\end{eqnarray}

\subsection{Power spectrum of $\delta \phi^I$ at the end of inflation}

To leading order, the two-point function of $\delta \phi^I$ at the end
of inflation is given by
\begin{eqnarray}
P^{IJ}~(N_f) =\Lambda^I_K (N_f,N_*) \Lambda^J_L (N_f,N_*) P^{KL}_\ast (k_1). \label{pm16}
\end{eqnarray}
Here, 
slightly after the horizon crossing,
$P^{IJ}_*$ is given to leading order by,
\begin{eqnarray}
P^{IJ}_* =\frac{ {(2\pi)}^3 }{4\pi k^3} { \left( \frac{H_*}{2\pi} \right) }^2 \delta^{IJ}\equiv P_* \delta^{IJ}. \label{pm17}
\end{eqnarray}
Any corrections to Eq.~(\ref{pm17}) are suppressed by the slow-roll parameters
\cite{Byrnes:2006vq}.  
Only fluctuations that extend to super-horizon scales at the end 
inflation contribute to the fluctuation of $\Gamma$. 
Such fields must be massless during inflation and we assume
$\Lambda^I_J \approx \delta^I_J$, which yields the corresponding power
spectrum as
\begin{eqnarray}
P^{IJ}(N_f)\approx P_* \delta^{IJ}. \label{pm18}
\end{eqnarray}

\subsection{Bispectrum of $\delta \phi^I$ at the end of inflation}

\begin{figure}[t]
  \includegraphics[width=9cm,clip]{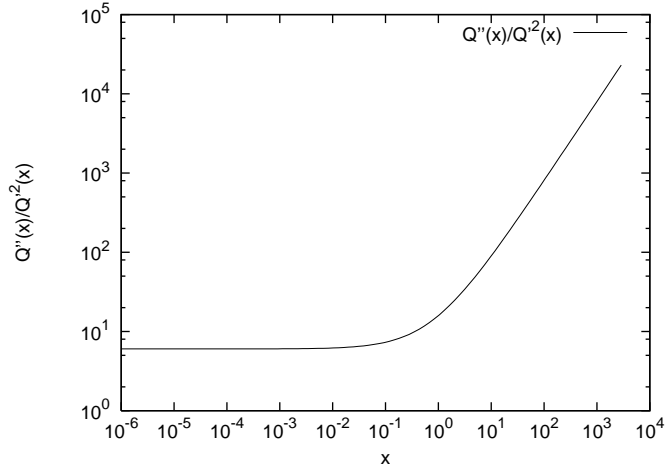}
\caption{Plot of $Q''(x)/Q'^2(x)$. 
}
\label{fig2} 
\end{figure}

To leading order, the bispectrum of $\delta \phi^I$ at the end of inflation
is given by
\begin{eqnarray}
B^{IJK}(k_1,k_2,k_3)(N_f) \simeq B^{IJK}_*(k_1,k_2,k_3)+\Theta^{IJK}(N_f,N_*) \left( P_\ast (k_1) P_\ast (k_2) +2~{\rm perms.} \right). \label{bi1}
\end{eqnarray}
$B^{IJK}_*$, 
which is the bispectrum slightly after the horizon crossing, 
can be evaluated by quantum perturbation theory \cite{zal}
\begin{eqnarray}
B^{IJK}_*(k_1,k_2,k_3)=-2 V_{IJK} {\rm Re} \bigg[ ig_{k_1} (\eta_*) g_{k_2} (\eta_*) g_{k_3} (\eta_*) \int_{-\infty}^{\eta_*} d\eta~a^4(\eta) g^*_{k_1} (\eta) g^*_{k_2} (\eta) g^*_{k_3} (\eta) \bigg], \label{bi2} 
\end{eqnarray}
where $g_k (\eta)$ is the mode function in de-Sitter spacetime and 
is given by,
\begin{eqnarray}
g_k (\eta)=\frac{iH}{\sqrt{2} k^{3/2}} (1+ik\eta ) e^{-ik\eta}.
\end{eqnarray}
Soon after the horizon crossing,
$k_i \eta$ becomes smaller than unity.
In such a phase,
the leading bispectrum of Eq.~(\ref{bi2}) is given by
\begin{eqnarray}
&&B_*^{IJK}(k_1,k_2,k_3)=-\frac{H_*^2}{4} \frac{k_t^3}{ {(k_1k_2k_3)}^3 } V_{IJK} \nonumber \\
&&\hspace{20mm}\times \Bigg[ \left( \gamma+\frac{1}{2} \log {(k_t \eta_* )}^2 \right) \left( -\frac{1}{3} +\frac{\sum_{i<j}k_ik_j}{k_t^2}-\frac{k_1k_2k_3}{k_t^3} \right)+\frac{4}{9}-\frac{\sum_{i<j}k_ik_j}{k_t^2} \Bigg], \label{bi3}
\end{eqnarray}
where $\gamma=0.577...$ is Euler's constant. \footnote{This expression 
for the bispectrum is slightly different from that given in Ref.~\cite{zal}, 
though the difference is not essential. 
Thanks to a second look by the author of Ref.~\cite{zal}, 
we reach the same result,
which is given above.}
Because the second line in Eq.~(\ref{bi3}) is ${\cal O}(1)$,
$B_*={\cal O}(V_{IJK}/H_*^4)$.
On the other hand,
the bispectrum of the second term in Eq.~(\ref{bi1}) can be estimated as
\begin{eqnarray}
\Theta^{IJK}(N_f,N_*)P_\ast (k_1) P_\ast (k_2)={\cal O} \left( \frac{1}{H_*^2} \int_{N_*}^{N_f} dN \frac{V_{IJK}}{H^2(N)} \right), \label{bi4}
\end{eqnarray}
which is expected to be larger than $B_*$ because $N_f - N_*$ is
typically $\CO(10)$. 
Hence the leading bispectrum of $\delta \phi^I$ at
the end of inflation comes from super-horizon evolution,
\begin{eqnarray}
B^{IJK}(N_f) \approx \Theta^{IJK}(N_f,N_*) \left( P_\ast (k_1) P_\ast (k_2)+2~{\rm perms.} \right). \label{bi5}
\end{eqnarray}
Thus, the corresponding non-linear parameter $f_{NL}$
becomes\footnote{Note that the definition of $f_{NL}$ here is different
in sign from that in \cite{Vernizzi:2003vs}}
\begin{eqnarray}
\frac{6}{5}f_{NL}=\frac{1}{xQ'(x)} \frac{\Gamma \Gamma_I {\tilde \Theta}^I}{\Gamma_K \Gamma^K}+\frac{1}{xQ'(x)} \frac{\Gamma \Gamma_I {\tilde \Gamma}^I_{(2)}}{{(\Gamma_K \Gamma^K)}^{3/2}}+\frac{Q''(x)}{Q'^2 (x)}, \label{bi6}
\end{eqnarray}
where ${\tilde \Theta}^I$ and ${\tilde \Gamma}_{(2)}^I$
are projected vectors of $\Theta^{IJK}$ and $\Gamma^{IJ}$,
respectively,
\begin{eqnarray}
{\tilde \Theta}^I =\frac{\Gamma_J \Gamma_K}{\Gamma_L \Gamma^L}\Theta^{IJK}, \label{bi7}\\
{\tilde \Gamma}_{(2)}^I =\frac{\Gamma_J \Gamma^{IJ}}{ {(\Gamma_K \Gamma^K)}^{1/2} }. \label{bi8}
\end{eqnarray}
One should notice that the first term in Eq.~(\ref{bi6}) represents intrinsic
non-Gaussianity due to cubic interactions of the scalar fields.  
The second term comes from non-linearity between $\Gamma$ and $\phi^I$.  
The third term comes from non-linearity between $\zeta$ and $\delta \Gamma$,
which only depends on $\Gamma$ through the argument $x = \Gamma/H_0$ of
$Q(x)$. 
$Q''(x)/Q'^2(x)$ takes minimum value $6$ at $x=0$ and becomes
larger for $x>1$ (see Fig.~\ref{fig2}). 
Hence if both the intrinsic non-Gaussianity of the scalar field fluctuations 
and the non-linearity between $\Gamma$ and $\phi^I$ are negligibly small, 
then $f_{NL}>5$.

\subsection{Trispectrum of $\delta \phi^I$ at the end of inflation}

\begin{figure}[t]
  \includegraphics[width=9cm,clip]{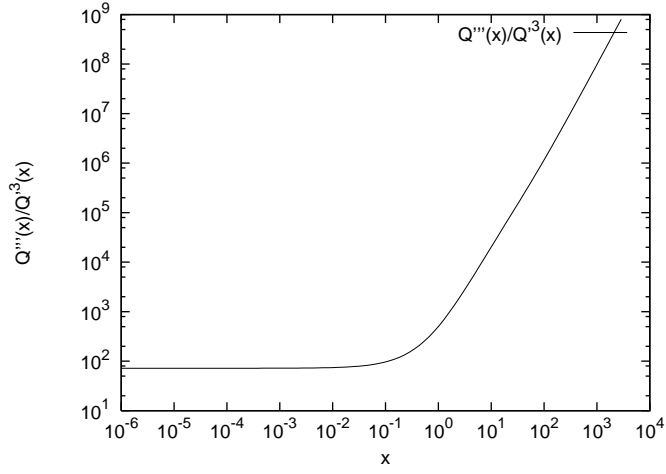}
\caption{Plot of $Q'''(x)/Q'^3(x)$. 
}
\label{fig3} 
\end{figure}

To leading order, the trispectrum of $\delta \phi^I$ at the end of
inflation is given by
\begin{eqnarray}
&&T^{IJKL}(k_1,k_2,k_3,k_4) (N_f)=T^{IJKL}_*(k_1,k_2,k_3,k_4) \nonumber \\
&&\hspace{10mm}+ \Theta^K_{IM}(N_f,N_*) \Theta^L_{JM}(N_f,N_*) \left( P_* (k_1) P_* (k_2) P_* (k_{13})+11~{\rm perms.} \right) \nonumber \\
&&\hspace{10mm}+\Xi^L_{IJK}(N_f,N_*) \left( P_* (k_1) P_* (k_2) P_* (k_3) +3~{\rm perms.} \right), \label{tri1}
\end{eqnarray}
As in the case of the bispectrum,
the leading contribution to $T^{IJKL}_*$ can be evaluated as \cite{zal}
\begin{eqnarray}
&&T^{IJKL}_*(k_1,k_2,k_3,k_4)=-2V_{IJKL} \nonumber \\
&&\hspace{15mm} \times {\rm Re} \Bigg[ ig_{k_1} (\eta_*)g_{k_2} (\eta_*)g_{k_3} (\eta_*)g_{k_4} (\eta_*) \int_{-\infty}^{\eta_*} d\eta~a^4(\eta) g^*_{k_1}(\eta)g^*_{k_2}(\eta)g^*_{k_3}(\eta)g^*_{k_4}(\eta) \Bigg]. \label{tri2}
\end{eqnarray}
For $k_i \eta_* \ll 1$,
Eq.~(\ref{tri2}) reduces to
\begin{eqnarray}
&&T^{IJKL}_*(k_1,k_2,k_3,k_4)=-\frac{H^4 k_t^3}{8 {(k_1k_2k_3k_4)}^3 } V_{IJKL} \nonumber \\
&&\hspace{40mm}\times \Bigg[ \left( \frac{1}{2} \log {(k_t \eta_*)}^2+\gamma \right) \left( -\frac{1}{3}+\frac{ \sum_{i<j} k_ik_j}{k_t^2} -\frac{ \sum_{i<j<\ell}k_ik_jk_\ell}{k_t^3} \right) \nonumber \\
&&\hspace{40mm}+\frac{4}{9}-\frac{ \sum_{i<j} k_ik_j}{k_t^2}+\frac{k_1k_2k_3k_4}{k_t^4} \Bigg], \label{tri3}
\end{eqnarray}
where $k_t=k_1+k_2+k_3+k_4$.  $T_*$ is typically smaller than the other
terms on the right hand side of Eq.~(\ref{tri1}), which corresponds to
the super-horizon evolution of the trispectrum due to the fourth order
interactions.  
Hence, as in the case of the bispectrum, 
the leading trispectrum of $\delta \phi^I$ at the end of inflation comes
from super-horizon evolution,
\begin{eqnarray}
&&T^{IJKL}(k_1,k_2,k_3,k_4) (N_f)=\Theta^K_{IM}(N_f,N_*) \Theta^L_{JM}(N_f,N_*) \left( P_* (k_1) P_* (k_2) P_* (k_{13})+11~{\rm perms.} \right) \nonumber \\
&&\hspace{50mm}+\Xi^L_{IJK}(N_f,N_*) \left( P_* (k_1) P_* (k_2) P_* (k_3) +3~{\rm perms.} \right), \label{tri4}
\end{eqnarray}
Therefore, the corresponding non-linear parameters $\tau_{NL}$ and
$g_{NL}$ are given by
\begin{eqnarray}
&&\tau_{NL}=\frac{1}{{(xQ'(x))}^2} \frac{\Gamma^2 {\tilde \Theta}^I {\tilde \Theta}_I }{\Gamma_K \Gamma^K}+\frac{2\Gamma^2 {\tilde \Gamma}^I_{(2)} {\tilde \Theta}_I}{ {(xQ'(x))}^2 {(\Gamma_K \Gamma^K)}^{3/2} }+\frac{2x^2 Q''(x) \Gamma \Gamma_I {\tilde \Theta}^I}{ {(xQ'(x))}^3 \Gamma_K \Gamma^K}\nonumber \\
&&\hspace{20mm}+\frac{\Gamma^2 {\tilde \Gamma}^I_{(2)} {\tilde \Gamma}_{I(2)} }{ {(xQ'(x))}^2 {(\Gamma_L \Gamma^L)}^2}+\frac{2x^2 Q''(x)}{ {(xQ'(x))}^3 } \frac{\Gamma \Gamma_I {\tilde \Gamma}^I_{(2)}}{ {(\Gamma_K \Gamma^K)}^{3/2} }+\frac{Q''^2(x)}{ Q'^4(x) }, \label{tri5} \\
&&\frac{54}{25}g_{NL}=\frac{\Gamma^2 {\tilde \Xi}}{ {(xQ'(x))}^2 \Gamma_I \Gamma^I}+\frac{1}{ {(xQ'(x))}^2} \frac{\Gamma^2 {\tilde \Gamma}^I_{(2)} {\tilde \Theta}_I}{ {(\Gamma_J \Gamma^J)}^{3/2} }+\frac{x^2 Q''(x)}{ {(xQ'(x))}^3 } \frac{\Gamma \Gamma_I {\tilde \Theta}^I}{\Gamma_K \Gamma^K} \nonumber \\
&&\hspace{20mm}+\frac{1}{ {(xQ'(x))}^2}\frac{\Gamma^2 \Gamma_I {\tilde \Gamma}^I_{(3)}}{ {(\Gamma_J \Gamma^J)}^2}+\frac{3x^2 Q''(x)}{ {(xQ'(x))}^3} \frac{\Gamma {\tilde \Gamma}^I_{(2)} \Gamma_I}{ {(\Gamma_J \Gamma^J)}^{3/2}}+\frac{Q'''(x)}{Q'^3(x)}, \label{tri6}
\end{eqnarray}
where
\begin{eqnarray}
{\tilde \Xi}=\frac{\Gamma_I \Gamma_J \Gamma_K \Gamma_L}{{(\Gamma_M \Gamma^M)}^2} \Xi^{IJKL},~~~~~{\tilde \Gamma}_{(3)}^I=\frac{\Gamma_J \Gamma_K \Gamma^{IJK}}{\Gamma_L \Gamma^L}. \label{tri7}
\end{eqnarray}
As in the case of the bispectrum, the last term of $g_{NL}$ represents
the contribution from the non-linear relation between $\zeta$ and
$\delta \Gamma$ and depends on $\Gamma$ only through the argument $x$ of
$Q(x)$.  
$Q'''(x)/Q'^3(x)$ takes minimum value $72$ at $x=0$, 
and larger values for $x>1$ (see Fig.~\ref{fig3}). 
We see that all three non-linear parameters, 
$f_{NL},~\tau_{NL}$ and $g_{NL}$, 
can be written only in terms of the vector quantities 
$\Gamma^I,~{\tilde \Gamma}^I_{(2)},~{\tilde \Gamma}^I_{(3)}$ 
and ${\tilde \Theta}^I$.

If $\Gamma^{IJK}$ and the intrinsic non-Gaussianity of scalar field 
fluctuations are negligibly small, 
we have the following relation
\begin{eqnarray}
g_{NL}=\frac{5}{3}\frac{Q''(x)}{Q'^2(x)}f_{NL}+\frac{25}{54}
 \frac{Q'''(x)}{Q'^3(x)}.
\label{tri8}
\end{eqnarray}
In particular,
for $x\ll 1$,
this relation reduces to
\begin{eqnarray}
g_{NL}=10f_{NL}-\frac{50}{3}. \label{tri9}
\end{eqnarray}
Hence $g_{NL}$ has the same order of magnitude as
$f_{NL}$ for $x\ll 1$ and becomes much larger than
$f_{NL}$ for $x \gg 1$.
This is in contrast with the case of standard single-field inflation,
where $g_{NL}$ is to second order in the slow-roll parameters.
Notice that the relation Eq.~(\ref{tri8}) depends on $\Gamma$ only
through the argument $x=\Gamma/H_0$ and is independent of the functional 
form of $\Gamma(\phi)$ except for conditions on $\Gamma^{IJK}$.

\section{Conclusions}

The modulated reheating scenario generates a primordial curvature
perturbation due to the spatial fluctuations of the inflaton decay rate.
Such a scenario induces larger non-Gaussianity in the perturbation than
that of the simple inflation scenario. 
Future observations such as the Planck satellite \cite{KK} may be able 
to detect or constrain the level of the non-Gaussianity, 
and thereby distinguish the different scenarios.

We have given expressions for the power spectrum, bispectrum and
trispectrum to leading order in the modulated reheating scenario,
allowing for multi-field dependence on the inflaton decay rate.  
The leading contribution to the bispectrum and trispectrum comes from the
super-horizon evolution.  
If the intrinsic non-Gaussianity of the scalar fields and third derivative 
of the decay rate $\Gamma^{IJK}$ are subdominant, 
then we have a simple relation between two non-linear parameters 
$f_{NL}$ and $g_{NL}$, which is independent of the detailed
form of the decay rate except for the condition on $\Gamma^{IJK}$.
Hence,
$g_{NL}$ has at least the same order of magnitude as $f_{NL}$.

We have also given a general inequality between the bispectrum
and the trispectrum $\tau_{NL} \ge \frac{36}{25}f_{NL}^2$ which 
is true for other inflationary scenarios as long as the non-Gaussianity
comes from the super-horizon evolution. 
This inequality allows $f_{NL}$ to be vanishingly small while $\tau_{NL}$ 
remains finite.
In such a case,
the leading non-Gaussianity comes not from the bispectrum
but from the trispectrum.

\section*{Acknowledgments}

We would like to thank Matias Zaldarriaga for useful comments.
We are also grateful to W.~Kelly for correcting the English.  
M.Y. is supported in part by JSPS Grant-in-Aid for Scientific Research
Nos.~18740157 and 19340054.

\end{document}